\documentclass[14pt]{article}
\usepackage{cite}
\usepackage{amssymb}
\usepackage{amsfonts}
\usepackage{mathrsfs}
\usepackage{graphicx}
\usepackage{amsmath}
\usepackage{color}
\usepackage{amsthm}
\usepackage{epstopdf}
\usepackage{booktabs}
\usepackage{pifont,bm}
\usepackage{tikz}
\usepackage{enumerate}
\usepackage{threeparttable}
\usepackage{epsfig}
\usepackage{subfigure}
\usepackage{float}
\usepackage{appendix}
\usepackage{relsize}
\usepackage{setspace}
\usepackage{multirow}

\theoremstyle{definition}

\makeatletter
\def\@biblabel#1{[#1]}
\makeatother

\makeatletter \@addtoreset{equation}{section}
\renewcommand{\theequation}{\arabic{section}.\arabic{equation}}

\begin{document}

\begin{titlepage}
\title{\bf { Rogue wave statistics and integrable turbulence in
the Gerdjikov-Ivanov equation
\footnotetext{
$\qquad^{\ast}$Corresponding author:
wqpeng@ouc.edu.cn; mathpeng@163.com (W.Q. Peng)\protect\\
$\qquad^{\dag}$ Corresponding author:
sftian@cumt.edu.cn; shoufu2006@126.com (S.F. Tian)}
}}

%
%
%

\author{Wei-Qi Peng$^{1,\ast}$, Xiao-Wang Lan$^{1}$, Shou-Fu Tian$^{2,\dag}$\\
\small \emph{$^{a}$School of Mathematical Sciences, Ocean University of China, Qingdao, 266100, China} \\
\small \emph{$^{b}$School of Mathematics, China University of Mining and Technology, Xuzhou 221116, China.} \\
\date{}}
\thispagestyle{empty}
\end{titlepage}
\maketitle

\vspace{-0.5cm}
\begin{center}
\rule{15cm}{1pt}\vspace{0.3cm}

\parbox{15cm}{\normalsize
{\bf Abstract}\\
\hspace{0.5cm}
This paper numerically investigates the statistical properties of rogue waves and their generation mechanisms in integrable turbulence, taking the Gerdjikov-Ivanov (GI) equation as the research object. First, based on the Lax pair of the GI equation, the eigenvalue spectra of the analytical solutions and the chaotic wave field are calculated using the Fourier collocation method. It is found that the spectral distribution of the GI equation exhibits a special symmetry with respect to both the real and imaginary axes. Subsequently, taking a plane wave with random noise as the initial condition, the evolution of chaotic wave fields is simulated using the split-step Fourier (SSF) method. Numerical results show that the larger the initial disturbance intensity, the faster the wave field converges to a chaotic state, and the higher the peak amplitude after convergence, the higher the tail of the probability density function, and the significantly higher probability of rogue wave occurrence. Moreover, as the initial disturbance intensity increases, the turbulence type transitions from breather turbulence to soliton turbulence. The increase in correlation length will mainly accelerate the convergence speed, have relatively little impact on peak amplitude and rogue wave probability, but will reduce the number of solitons in the chaotic field. Therefore, the increase in correlation length will not change the type of turbulence. In addition, the evolution of the wave-action spectrum is studied. The research has found that the wave-action spectrum of the GI equation shows an asymmetric distribution during the time evolution process, and this asymmetry persists even after the system reaches a steady state. The wave-action spectrum tends to stabilize relatively quickly. Its value is proportional to the amplitude of the random parameters and inversely proportional to the correlation length. This study reveals the basic characteristics of integrable turbulence in the GI equation and the conditions for rogue wave generation from numerical and statistical perspectives, providing a reference for understanding extreme wave phenomena in this type of integrable system.
}

\vspace{0.5cm}
\parbox{15cm}{\small{

\vspace{0.3cm} \emph{} \\

} }
\end{center}
\vspace{0.3cm} \rule{15cm}{1pt} \vspace{0.2cm}

\section{Introduction}\label{sec-1}
Turbulence is a complex and ubiquitous phenomenon characterized by multi-scale, irregular and chaotic motions. Usually, this phenomenon arises from the complex interactions among numerous nonlinear coherent structures, which give rise to the entire system to exhibit complex and seemingly unpredictable behaviors. Traditional studies on turbulence are mainly based on the Navier-Stokes equations. In recent years, however, complex turbulence-like behaviors have also been discovered in integrable systems, which are defined as integrable turbulence \cite{5,6}.
In the study of integrable turbulence, researchers have discovered a common yet peculiar phenomenon: giant waves with peak amplitudes several times larger than the surrounding waves suddenly appear at extremely high speeds and then disappear within a very short period of time. Such extreme phenomena are called ``rogue waves''. The concept of rogue waves was first proposed by oceanographers to describe abnormal ocean waves with extreme amplitudes that occur suddenly in marine environments \cite{8,9}. Subsequently, analogous extreme events have been observed in various physical systems, including nonlinear optics, plasmas and superfluids \cite{10,11}. The rogue waves have distinct local characteristics, and their formation and disappearance occur extremely suddenly. Their peak amplitudes can reach two to three times or even higher than those of background waves \cite{12}. From a mathematical perspective, rogue waves correspond to a class of rational solutions to integrable equations. Among them, the Peregrine soliton of the nonlinear Schr\"{o}dinger (NLS) equation is the most representative \cite{13}. In realistic physical scenarios, the background of waves is usually dominated by random fluctuations or turbulence caused by modulation instability. Rogue waves essentially emerge as statistical extreme values resulting from the interaction of a large number of nonlinear coherent structures \cite{14}. Therefore, in order to clarify the statistical characteristics of the rogue waves (such as their occurrence probability in the turbulent background, amplitude distribution and spectral features), it is necessary to combine the analytical exact solutions with the evolution theory of integrable turbulence.

Remarkable advances in the theoretical analysis and numerical simulation of integrable turbulence have been achieved in recent years. In 2009, Zakharov established a systematic theoretical framework for turbulence in integrable systems and proved that the energy transfer in integrable turbulence is subject to strict conservation laws, where solitons and breathing solitons are the basic excitation units \cite{5}. Agafontsev and Zakharov used numerical methods to study the integrable turbulence caused by modulation instability (MI) in the NLS equation, and confirmed that the intensity of the initial random perturbation has a significant impact on the probability of rogue waves generation \cite{6}. Soto-Crespo et al. \cite{15} clearly distinguished between breather turbulence and soliton turbulence for the first time, and revealed the transition mechanism of the two turbulent states via eigenvalue distribution derived from the inverse scattering transform. Specifically, a weak initial perturbation will cause the eigenvalues to concentrate on the imaginary axis, which corresponds to the formation of Akhmediev breathing modes (i.e., breathing turbulence). As the intensity of the initial perturbation increases, the eigenvalues gradually deviate from the imaginary axis, resulting in solitons with non-zero real parts, and eventually leading to a transition to soliton turbulence. They also explained that when the initial state is a random function with a sufficiently high amplitude, the probability of the occurrence of rogue waves in such chaotic wave states significantly increases. Roberti et al. investigated the statistical characteristics of integrable turbulence for the one-dimensional small-dispersion NLS equation, and by utilizing the scale separation in the semiclassical region, they obtained a simple explicit formula that describes the evolution of the fourth-order moment of the random wave field amplitude in the early stage \cite{jia12}.
 Sun et al. reported the numerical experiments of turbulence and rogue waves for two nearly-integrable NLS equations with the effect of disorder, and they discovered that by adding a weak system noise, the probability of rogue waves occurrence could be significantly increased \cite{jia11}.  Congy et al. utilized the spectral dynamics theory of soliton gases to study the probability of occurrence of extreme events in the integrable turbulence described by the focusing NLS equation, providing a theoretical explanation for the statistical characteristics of rogue waves that have long puzzled the academic community \cite{jia3}.
On the experimental front, Suret et al. were the first to employ the delayed microscopy technique to capture the evolution process of integrable turbulence in optical fibers in real time, and they experimentally verified the generation of Peregrine solitons \cite{17}. Michel et al. observed the early-stage evolution of integrable turbulence in unidirectional deep-water gravity wave experiments \cite{18}. These experimental observations are in excellent agreement with numerical simulations, providing strong evidence to validate the theoretical framework of integrable turbulence.

Furthermore, the research on integrable turbulence has continued to expand to various integrable models. Didenkul conducted a numerical simulation of soliton turbulence for the focused Gardner equation. The results showed that the interaction between opposite-polarity solitary waves is the core mechanism for generating large-amplitude rogue waves.\cite{19}. Wang et al. studied integrable turbulence in the Hirota equation and analyzed the variation of eigenvalue distributions under different parametric conditions \cite{20}. Xie et al. focused on the inherent properties of integrable turbulence in coupled NLS system \cite{21} and coupled cubic-quintic NLS system \cite{jia9}. Tang et al. focus on the fourth-order NLS equation and analyse the dynamics of integrable turbulence \cite{jia10}. Xu et al. conducted a systematic numerical study on the interaction dynamics of solitary waves and the statistical characteristics of solitary wave turbulence based on the Gardner equation \cite{22,22.5}. They particularly focused on the influence of the interaction between opposite-polarity solitary waves on skewness and kurtosis. From the perspective of eigenvalue distribution, Li et al. investigated integrable turbulence in the Kundu-Eckhaus equation, and found that high-order nonlinear terms can significantly enhance the generation probability of rogue waves \cite{23}. The integrable turbulence problem of the modified NLS equation with self-steepening effect has also been studied in Ref. \cite{jia4}.  Yan et al. conducted a comprehensive analysis of the integrable turbulence and rogue waves generated by the MI of plane waves in the derivative nonlinear Schr\"{o}dinger (DNLS) equation \cite{43} and the fractional NLS equation \cite{43.5}, respectively.

Inspired by the above studies, a critical research issue arises naturally: for integrable models with complex nonlinear terms (such as derivative-type nonlinearity and quintic nonlinearity) represented by the Gerdjikov-Ivanov (GI) equation, what are the statistical characteristics of their integrable turbulence and the dynamic mechanism underlying the transition from breathers to solitons? In addition, existing studies mainly focus on the evolution of statistical moments in soliton turbulence, while the evolution of the wave action spectrum and its correlation with initial random parameters remain insufficiently explored.

The Ablowitz-Kaup-Newell-Segur (AKNS) system \cite{28} is one of the very important integrable systems related to linear spectral problems, which contains the well-known NLS equation \cite{29}. Based on this linear spectrum, it can be generalized to any-order polynomial spectral problems. \cite{30}. It should be noted that the Kirillov-Kostant form corresponding to the general polynomial spectral problem is degenerate,  which imposes specific constraints on the integrable partial differential equations. Under distinct constraint conditions, the quadratic polynomial spectral problems correspond to three types of DNLS equations, namely the DNLSI, DNLSII and DNLSIII equations. Their explicit expressions are presented in Ref. \cite{1}.
\\
The mathematical form of the DNLSI equation is written as:\begin{align}\label{1}
u_{t}+iu_{xx}+(|u|^{2}u)_{x}=0.
\end{align}
\\
The DNLSII equation is:
\begin{align}\label{2}
iu_{t}+u_{xx}+iuu^{\ast}u_{x}=0.
\end{align}
\\
The DNLSIII  takes the form:
\begin{align}\label{3}
iu_{t}+u_{xx}-iu^{2}u_{x}^{\ast}+\frac{1}{2}u^{3}(u^{\ast})^{2}=0,
\end{align}
where the asterisk denotes the complex conjugate, and the subscript represents the partial derivative with respect to
$x$ or $t$. Specifically, this equation was first proposed by Gerdjikov and Ivanov in Ref. \cite{32}, and is also widely known as the GI equation. In physical contexts, $u(x,t)$ generally describes the perturbation function of the transverse magnetic field, and is often used to characterize the Alfv\'{e}n waves propagating along the direction of the ambient magnetic field in plasma physics. Over the past few decades, extensive efforts have been devoted to investigating various analytical solutions and critical properties of the GI equation via diverse methods. Using the Darboux transformation (DT) method, researchers have derived breather and rogue wave solutions of the GI equation \cite{1,31,jia5,jia8}. Breathers and rogue waves on the periodic background for the GI equation have been also derived in Ref. \cite{jia6}. The soliton molecules and
dynamics of the smooth positons for the GI equation were discussed in Ref. \cite{jia7}. The long-time asymptotic behavior for GI equation has
been analysed \cite{33,34,35}. Meanwhile, its soliton solutions have been obtained systematically through the Riemann-Hilbert approach \cite{36,37,36.5,37.5}. The data-driven localized solutions and inverse problems of the GI equation were investigated by employing physics-informed neural network (PINN) algorithm \cite{37.51}.

Although a large number of studies have focused on the GI equation, there is still a lack of systematic numerical research on the statistical evolution characteristics of this equation under random initial conditions and the spectral properties of rogue wave solutions. In particular, two key issues remain insufficiently explored: How to understand the relationship between rogue waves and background turbulence from the spectral domain perspective, and how random wave fields generate observable extreme events through mechanisms such as modulation instability.
Against this background, this paper adopts efficient numerical methods to systematically investigate the rogue wave statistics and turbulence phenomena described by the GI equation. The main contributions are summarized as follows:
\begin{enumerate}
\item This work innovatively addresses the quadratic eigenvalue problem of spectral parameters corresponding to the Lax pair of the GI equation and obtains the eigenvalue spectra. The results reveal that the eigenvalue spectral structure of analytical solutions exhibits a symmetric distribution with respect to two coordinate axes. We further calculate the spectra under random initial potentials and analyze the types of solutions in integrable turbulence, along with the effects of random parameters on turbulent solution modes.
\item The long-time evolution of the GI equation with random initial potential functions is analyzed in detail. By configuring random initial fields with different correlation lengths and amplitude distributions, the evolution process from initial noise to a fully turbulent state is numerically simulated. It is found that the intensity of random perturbation is positively correlated with the occurrence probability of rogue waves and dominates the classification of turbulence regimes. Meanwhile, numerical results indicate that the correlation length of random perturbations exerts no significant influence on rogue wave probability or turbulence type. However, it can effectively reduce the number of solitons in the wave field.
\item The wave action spectrum in turbulence is calculated, and its dynamic evolution as well as the modulation effects of random parameters are comprehensively discussed. A unique property of the GI equation is identified: its wave action spectrum presents an asymmetric distribution during temporal evolution. After reaching a steady state, the wave action spectrum is positively correlated with the intensity of the random field and negatively correlated with the correlation length.
\end{enumerate}

The outline of this article is structured as follows: In Section 2, we mainly present the analytical solutions of various types for the GI equation, and explore the influence of spectral parameters on the specific forms of these analytical solutions. Subsequently, the Fourier collocation method is applied to plot the eigenvalue spectra of the analytical solutions.  In Sections 3, the split-step Fourier method (SSF) method is adopted to sanalyzed the changing trends of various statistical indicators during the occurrence of turbulence, where a plane wave with random noise is set as the initial condition. We analyze the generation mechanism of rogue waves induced by the interaction between solitons and breathers in integrable turbulence. The conclusion will be presented in the final section.

\section{Analytical solutions and corresponding spectral characteristics of the GI equation}
In this section, we mainly present various analytical solutions of the GI equation and investigate the influence of spectral parameters on the classification of these solutions. Furthermore, the Fourier collocation method is applied to calculate the spectra corresponding to the analytical solutions of the GI equation, and the spectral characteristics of different types of analytical solutions are illustrated.
\subsection{Various analytical solutions}
The Lax pair of Eq. \eqref{3} can be readily derived from References \cite{1,31}, and its specific form is presented as follows:
\begin{align}\label{4}
&\partial_{x}\psi=\left(J\lambda^{2}+Q_{1}\lambda+Q_{0}\right)\psi=U\psi,\notag\\
&\partial_{t}\psi=\left(2J\lambda^{4}+V_{3}\lambda^{3}+V_{2}\lambda^{2}+V_{1}\lambda+V_{0}\right)\psi=V\psi,
\end{align}
where
\begin{align}\label{5}
&\psi=\left(\begin{array}{c}
    \phi  \\
    \varphi \\
\end{array}\right)=\left(\begin{array}{c}
    \phi(x,t,\lambda)  \\
    \varphi(x,t\lambda) \\
\end{array}\right), \qquad J=\left(\begin{array}{cc}
    -i  &  0\\
    0 &  i\\
\end{array}\right),\qquad Q_{1}=\left(\begin{array}{cc}
    0  &  u\\
    v &  0\\
\end{array}\right),\notag\\
&Q_{0}=\left(\begin{array}{cc}
    -\frac{1}{2}iuv  &  0\\
    0 &  \frac{1}{2}iuv\\
\end{array}\right),\qquad V_{3}=2Q_{1},\qquad V_{2}=Juv,\qquad V_{1}=\left(\begin{array}{cc}
    0  &  iu_{x}\\
    -iv_{x} &  0\\
\end{array}\right),\notag\\
&V_{0}=\left(\begin{array}{cc}
    \frac{1}{2}(vu_{x}-uv_{x})+\frac{1}{4}iu^{2}v^{2}  &  0\\
    0 &  -\frac{1}{2}(vu_{x}-uv_{x})-\frac{1}{4}iu^{2}v^{2}\\
\end{array}\right),
\end{align}
where $v=-u^{\ast}$. $\lambda\in\mathbb{C}$, $\psi$ denotes the eigenfunction of Eq. \eqref{4} with respect to the eigenvalue $\lambda$. Let $\lambda=\alpha+i\beta$.

The soliton, breather and rogue wave solutions of Eq. \eqref{3} can be obtained from Refs. \cite{1,31}, which are given below (where
$\alpha_{1}, \beta_{1}$ represent the real and imaginary parts of the eigenvalue, respectively):\\
A. Soliton solution
\begin{align}\label{6}
u=\frac{-8e^{F_{1}}\alpha_{1}\beta_{1}}{-e^{F_{2}}\alpha_{1}+ie^{F_{2}}\beta_{1}-e^{-F_{2}}\alpha_{1}-ie^{-F_{2}}\beta_{1}},
\end{align}
wher
\begin{align}\label{7}
&F_{1}=-2i\left(\alpha_{1}^{2}x+2\alpha_{1}^{4}t-12\alpha_{1}^{2}\beta_{1}^{2}t-\beta_{1}^{2}x+2\beta_{1}^{4}t\right),\notag\\
&F_{2}=4\alpha_{1}\beta_{1}\left(4t\alpha_{1}^{2}-4t\beta_{1}^{2}+x\right).
\end{align}
It is a line soliton, whose trajectory on the $(x,t)$ plane satisfies
\begin{align}\label{8}
x=-4t(\alpha_{1}-\beta_{1})(\alpha_{1}+\beta_{1}).
\end{align}
By taking the limit $\alpha_{1}\rightarrow0 $ in Eq. \eqref{6}, we obtain the rational traveling-wave soliton solution
\begin{align}\label{9}
u=\frac{4\beta_{1}e^{-2i\beta_{1}^{2}\left(-x+2t\beta_{1}^{2}\right)}}{1-4i\beta_{1}^{2}x+16i\beta_{1}^{4}t}.
\end{align}

\begin{figure}[H]
~~~~~~~~~~~~~~
\includegraphics[width=4.8cm,height=4.8cm,angle=0]{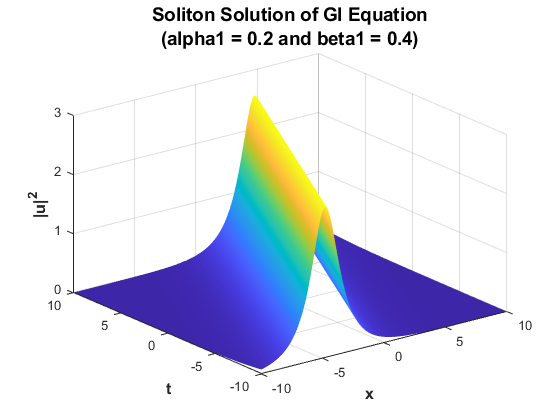}
~~~~~~~~
\includegraphics[width=4.8cm,height=4.8cm,angle=0]{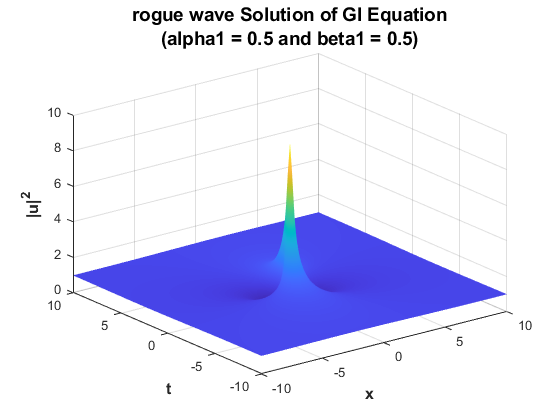}

~~~~~~~~\qquad\qquad\quad\qquad\qquad
$\textbf{(a)}\qquad\qquad\qquad\qquad\qquad\qquad\quad\qquad
 \textbf{(b)}$\\
 ~~~~~~
\includegraphics[width=4.8cm,height=4.8cm,angle=0]{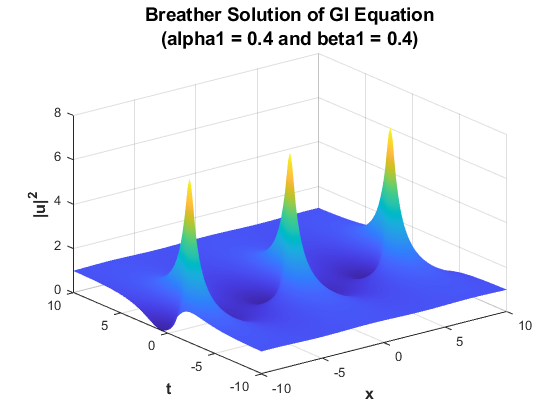}
\includegraphics[width=4.8cm,height=4.8cm,angle=0]{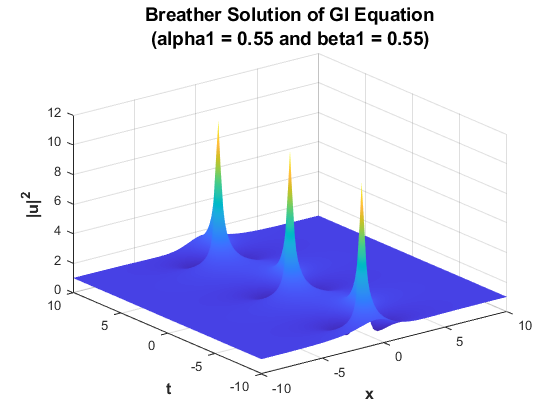}
\includegraphics[width=4.8cm,height=4.8cm,angle=0]{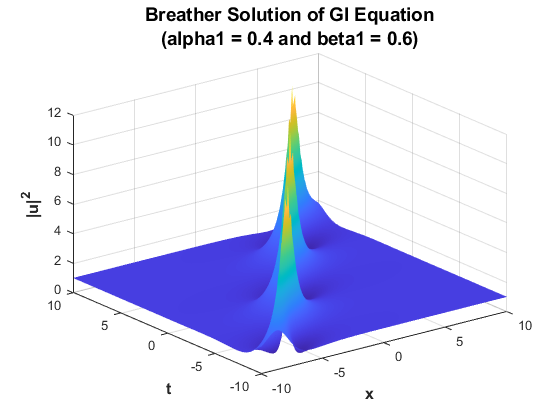}

~~~~~~~~\qquad
$\textbf{(c)}\qquad\qquad\qquad\qquad\qquad\qquad\qquad
 \textbf{(d)}\qquad\qquad\qquad\qquad\qquad\quad\qquad\textbf{(e)}$\\
\caption{\small Profiles of analytical solutions to Eq. \eqref{3}: (a) Soliton solution with $\alpha_{1}=0.2,\beta_{1}=0.4$; (b) Rogue wave solution with $\alpha_{1}=\beta_{1}=0.5$;(c) Spatially periodic breather solution with $\alpha_{1}=\beta_{1}=0.4$; (d) Temporally periodic breather solution with $\alpha_{1}=\beta_{1}=0.55$; (e) Spatiotemporally periodic breather solution with $\alpha_{1}=0.4, \beta_{1}=0.6$.}
\label{fig1}
\end{figure}

It can be seen from Eq. \eqref{8} that the soliton velocity is related to $(\alpha_{1}-\beta_{1})(\alpha_{1}+\beta_{1})$, namely, it depends on both spectral parameters. Moreover, as reported in Refs. \cite{1,31}, the soliton amplitude is only associated with the imaginary part of the spectral parameter.
By setting $\alpha_{1}=0.2,\beta_{1}=0.4$
in Eq. \eqref{6}, the evolution process of the soliton solution is obtained, as illustrated in Fig. \ref{fig1}(a). Substituting the same parameters into Eq. \eqref{8} yields the soliton trajectory $x=0.48t$ on the $(x,t)$ plane, which is in good agreement with the trajectory shown in the figure.
\\
B. Rogue wave solution
\begin{align}\label{10}
u=\frac{G_{1}}{G_{2}}\exp\left(\frac{1}{2}i\left(2ax-2ta^{2}-2tc^{2}a+tc^{4}\right)\right),
\end{align}
where
\begin{align}\label{11}
G_{1}=&-8c^{2}a^{3}t^{2}+12a^{2}c^{4}t^{2}+8c^{2}a^{2}tx-8c^{4}atx-2c^{2}ax^{2}\notag\\
&+12iac^{2}t-6ac^{6}
t^{2}-3-2ic^{2}x+2c^{4}x^{2}-6ic^{4}t+2c^{8}t^{2},\notag\\
G_{2}=&-8c^{2}a^{3}t^{2}+12a^{2}c^{4}t^{2}+8c^{2}a^{2}tx-8c^{4}tax-2c^{2}ax^{2}\notag\\
&+4iac^{2}t-6ac^{6}t^{2}+1-2ic^{2}x+2c^{4}x^{2}+2ic^{4}t+2c^{8}t^{2},
\end{align}
$a$ and $c$ are real parameters, which satisfy: $a=-2\alpha_{1}^{2}+2\beta_{1}^{2}$, $\mathrm{Im}(-c^{4}+2c^{2}a-a^{2}-4a\lambda_{1}^{2}-4\lambda_{1}^{4})=0$. By setting
$a=0$ and $c=1$(i.e., taking $\alpha_{1}=\beta_{1}=0.5$), we obtain the evolution of the rogue wave, as shown in Fig. \ref{fig1}(b).\\
C. Breather wave solution
\begin{align}\label{12}
u=\left(c+\frac{u_{1}}{u_{2}}\right)\exp\left(i\left(2xS+\left(-4S^{2}-2c^{2}S+\frac{1}{2}c^{4}\right)t\right)\right),
\end{align}
where
\begin{align}\label{13}
u_{1}=&4\alpha_{1}\beta_{1}\left(\left(c-2\beta_{1}\right)^{2}\left(c^{2}+4\alpha_{1}^{2}\right)
-K_{1}^{2}\right)\cosh\left(f_{1}\right)\notag\\
&+4\alpha_{1}\beta_{1}\left(\left(c-2\beta_{1}\right)^{2}\left(
c^{2}+4\alpha_{1}^{2}\right)+K_{1}^{2}\right)\cos\left(f_{2}\right)\notag\\
&-16i\alpha_{1}^{2}\beta_{1}\left(c-2\beta_{1}
\right)K_{1}\sin\left(f_{2}\right)-8ic\alpha_{1}\beta_{1}\left(c-2\beta_{1}\right)K_{1}\sinh\left(f_{1}\right),\notag\\
u_{2}=&\alpha_{1}\left(\left(c-2\beta_{1}\right)^{2}\left(c^{2}+4\alpha_{1}^{2}\right)+K_{1}^{2}\right)\cosh\left(f_{1}\right)\notag\\
&+\alpha_{1}\left(\left(c-2\beta_{1}\right)^{2}\left(c^{2}+4\alpha_{1}^{2}\right)-K_{1}^{2}\right)\cos\left(f_{2}\right)\notag\\
&-2ic\beta_{1}\left(c-2\beta_{1}\right)K_{1}\sin\left(f_{2}\right)+4i\alpha_{1}\beta_{1}\left(c-2\beta_{1}\right)K_{1}\sinh\left(f_{1}\right)
\end{align}
and
\begin{align}\label{14}
&S=\beta_{1}^{2}-\alpha_{1}^{2},\qquad f_{1}=K_{1}\left(4\left(\alpha_{1}^{2}-\beta_{1}^{2}\right)t+x\right),\notag\\ &f_{2}=4K_{1}\alpha_{1}\beta_{1}t,\quad K_{1}=\sqrt{16\alpha_{1}^{2}\beta_{1}^{2}-4c^{2}\alpha_{1}^{2}+4c^{2}\beta_{1}^{2}-c^{4}}.
\end{align}
For the breather solution, as $x\rightarrow\infty$ and $t\rightarrow\infty$, it follows that $\vert q\vert^{2}\rightarrow c^{2}$.
When $K_{1}^{2}>0$, setting $f_{1}=0$
 yields the motion trajectory of the solution as follows:
$
x=-4\alpha_{1}^{2}t+4\beta_{1}^{2}t.
$
When $K_{1}^{2}<0$, setting $f_{2}=0$ gives the corresponding motion trajectory of the solution:
$
t=0.
$

Thus, the type of breather (spatial breather or temporal breather) can be controlled by adjusting the magnitudes of the real and imaginary parts of the eigenvalue. For instance, under the condition $c=1$:

Setting $\alpha_{1}=\beta_{1}=0.4$
 yields $K_{1}^{2}<0$, corresponding to the spatially periodic breather (Akhmediev breather) illustrated in Fig. \ref{fig1}(c). Its trajectory is
$t=0$, which is consistent with the numerical results.

Setting $\alpha_{1}=\beta_{1}=0.55$
 yields $K_{1}^{2}>0$, corresponding to the temporally periodic breather (Kuznetsov-Ma breather) illustrated in Fig. \ref{fig1}(d). Since $\alpha_{1}=\beta_{1}$, the trajectory of the solution is $x=0$, which matches the figure.

Setting $\alpha_{1}=0.4,\beta_{1}=0.6$
 yields $K_{1}^{2}>0$, corresponding to the spatiotemporally periodic breather illustrated in Fig. \ref{fig1}(e). The trajectory of the solution is derived as
$x=0.8t$, which is in good agreement with the figure.

The explicit forms of the above three types of analytical solutions have been presented. We use MATLAB to plot the corresponding waveforms, and the results are displayed in Fig. \ref{fig1}.

\subsection{The associated spectra of the analytical solutions}
In the previous section, the analytical solutions and Lax pair of the GI equation have been presented. However, it is rather difficult to directly compute the spectra using the given Lax pair. Accordingly, this section introduces the Fourier collocation method for solving the corresponding spectra. To obtain the spectra with the analytical solutions taken as the potential function via the Fourier collocation method \cite{jia1}, we adopt the spatial component of the Lax pair \eqref{4} and reformulate it into the following form:
\begin{align}\label{17}
\left(J\lambda^{2}+Q_{1}\lambda+G\right)\psi=0,
\end{align}
where $J$ and $Q_1$ are given by Eq. \eqref{5}. Let \(G = Q_0 - \partial_x\), where \(Q_0\) is defined in Eq. \eqref{5}, and \(\partial_x\) denotes the partial derivative with respect to $x$. In this way, the problem of solving the spectrum is transformed into a quadratic eigenvalue problem. Nevertheless, for a general potential function \(u(x,t)\), it is difficult to analytically obtain the spectrum from the above equation.
In this work, we employ the Fourier collocation method to numerically solve Eq. \eqref{17} and acquire the corresponding spectral diagrams. First, the $x$-axis is truncated into a finite interval of length $L$, and the functions \(\phi(x)\), \(\varphi(x)\) and \(u(x,t)\) are expanded in Fourier series:
\begin{align}\label{18}
&\phi=\sum_{n=-N}^{N}a_{1n}e^{ink_{0x}},\qquad \varphi=\sum_{n=-N}^{N}a_{2n}e^{ink_{0x}},\notag\\ &u=\sum_{n=-N}^{N}b_{n}e^{ink_{0x}},\qquad u^{\ast}=\sum_{n=-N}^{N}b_{n}^{\ast}e^{ink_{0x}},\qquad uu^{\ast}=\sum_{n=-N}^{N}c_{n}e^{ink_{0x}},
\end{align}
where \(k_0=\dfrac{2\pi}{L}\) and \(t=t_0\). Substituting Eq. \eqref{18} into Eq. \eqref{17} and equating the corresponding Fourier coefficients, the associated eigenvalues (spectra) can be efficiently computed by employing the built-in MATLAB function \emph{polyeig} for solving quadratic eigenvalue problems.

\begin{figure}[H]
~~~~~~~~~~~~~~
\includegraphics[width=4.8cm,height=4.6cm,angle=0]{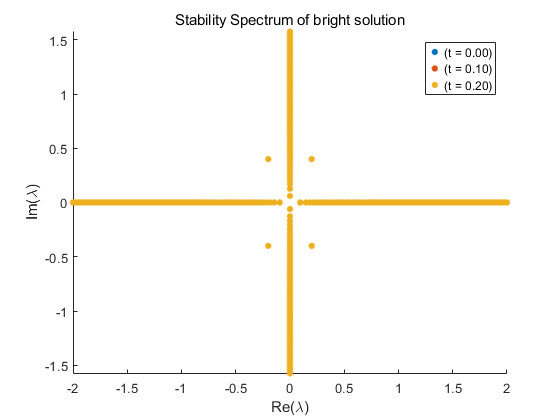}
~~~~~~~~
\includegraphics[width=4.8cm,height=4.6cm,angle=0]{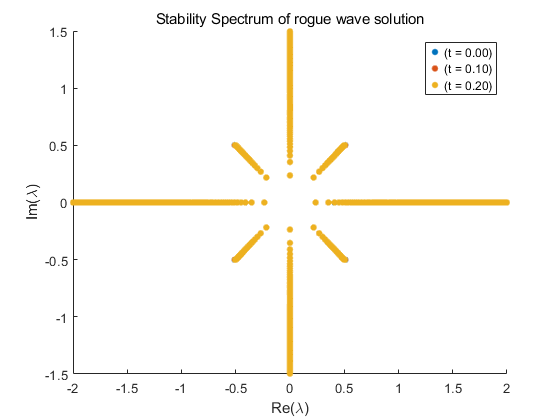}

~~~~~~~~\qquad\qquad\quad\qquad\qquad
$\textbf{(a)}\qquad\qquad\qquad\qquad\qquad\qquad\quad\qquad
 \textbf{(b)}$\\
 ~~~~~~
\includegraphics[width=4.8cm,height=4.6cm,angle=0]{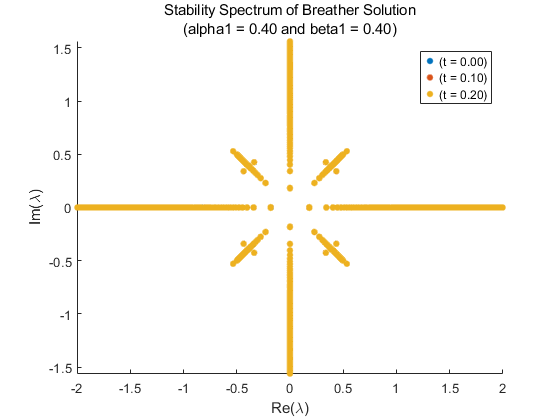}
\includegraphics[width=4.8cm,height=4.6cm,angle=0]{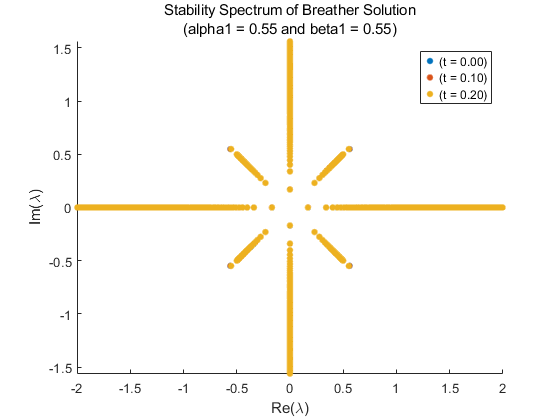}
\includegraphics[width=4.8cm,height=4.6cm,angle=0]{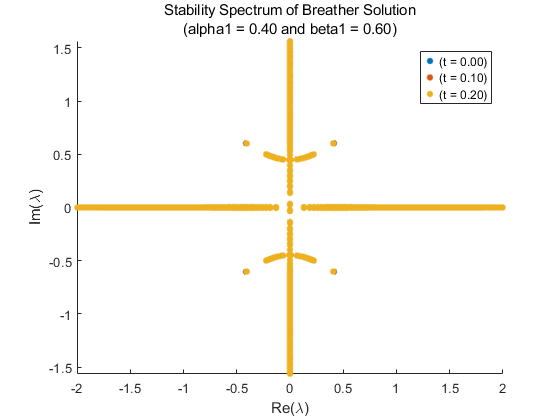}

~~~~~~~~\qquad
$\textbf{(c)}\qquad\qquad\qquad\qquad\qquad\qquad\qquad
 \textbf{(d)}\qquad\qquad\qquad\qquad\qquad\quad\qquad\textbf{(e)}$\\
\caption{\small(a)-(e) are the associated spectra of the analytical solutions of Eq. \eqref{3} corresponding to Fig. \ref{fig1}(a)-(e).}
\label{fig2}
\end{figure}

Fig. \ref{fig2}(a)-(e) present their corresponding spectral diagrams, which characterize the distribution of spectral parameters in the real and imaginary parts, denoted by \(\mathrm{Re}(\lambda)\) and \(\mathrm{Im}(\lambda)\), respectively. It can be observed from the spectral results that the eigenvalues are independent of time, which is consistent with the theoretical analysis. Furthermore, an interesting phenomenon is found that the spectra are symmetric with respect to both the real axis and the imaginary axis. That is, the eigenvalues appear in quadruple sets.

By analyzing the analytical solutions of breathers, it can be seen that the value of \(K_1\) is governed by \(\alpha_1^2\) and \(\beta_1^2\). This clearly implies that if \(\lambda=\alpha+i\beta\) is the spectral parameter of a spatial breather, then \(\lambda=\pm\alpha\pm i\beta\) are also spectral parameters of the breather, which is consistent with the plotted results.

A careful observation of the figures shows that Fig. \ref{fig2}(a) contains four symmetric points about the coordinate axes: \(\pm0.2\pm0.4i\), which include the preset parameter \(\lambda=0.2+0.4i\) of the soliton analytical solution. Similarly, Fig. \ref{fig2}(b) presents the axisymmetric points \(\pm0.5\pm0.5i\), corresponding to the preset parameter \(\lambda=0.5+0.5i\) of the analytical solution; Fig. \ref{fig2}(c) contains the symmetric points \(\pm0.4\pm0.4i\); Fig. \ref{fig2}(d) includes \(\pm0.55\pm0.55i\); and Fig. \ref{fig2}(e) possesses \(\pm0.4\pm0.6i\). All these cases exactly contain the spectral parameters adopted for plotting the analytical solution profiles.

\section{Numerical simulation of turbulence in the GI equation}
In this section, we aim to study the statistical properties of soliton and breather turbulence in the chaotic wave field, and then analyze the generation mechanism of rogue waves induced by the interaction between solitons and breathers in integrable turbulence.

To simulate the chaotic wave field, we adopt a plane wave with random noise as the initial condition:
\begin{align}\label{19}
u(x, 0) = 1 + \mu f(x)
\end{align}
where \(f(x)\) is a normalized complex random function with a standard deviation \(\sigma = 1\), and \(\mu\) denotes the standard deviation of the entire function \(u(x, 0)\) \cite{14,15,39}. Both the real and imaginary parts of \(f(x)\) are discretized numerically as the following Gaussian distribution:
\begin{align}\label{20}
\sqrt{\frac{2\Delta x}{\sqrt{\pi} L_c}} \exp\left[-2\left(\frac{x_{k}}{L_c}\right)^2\right]
\end{align}
where \(L_c\) represents the correlation length of the random values, a quantity that measures the critical threshold for the mutual independence between two random points. Since the initial condition is governed by \(\mu\) and \(L_c\), we investigate the effects of these two parameters on the integrable turbulence of Eq. \eqref{3}.

As the experiments are performed under random conditions, slight discrepancies exist in the results of each program run. However, the overall properties and phenomena remain consistent, as the inherent nature of the equation is unaffected by the initial randomness.  The robustness of the experimental results against random factors in the program can be verified by running the attached MATLAB code multiple times.

\subsection{Spectral characteristics of the chaotic wave field}
\begin{figure}[H]
 \includegraphics[width=4.8cm,height=4.6cm,angle=0]{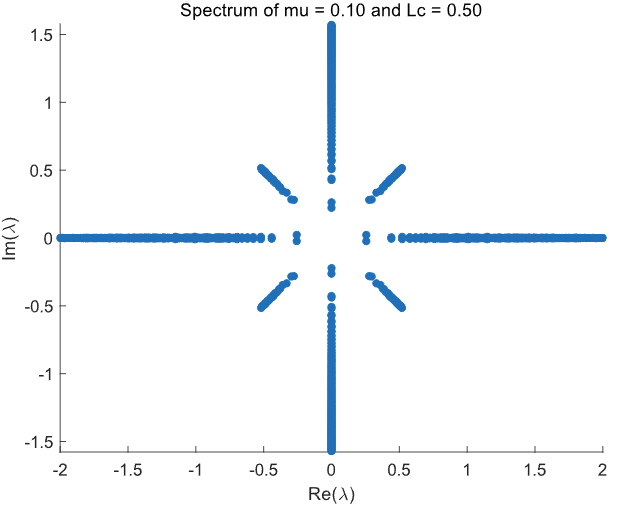}
\includegraphics[width=4.8cm,height=4.6cm,angle=0]{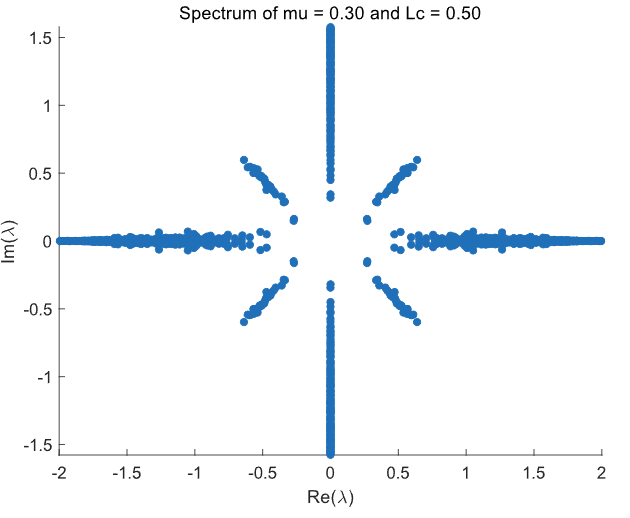}
\includegraphics[width=4.8cm,height=4.6cm,angle=0]{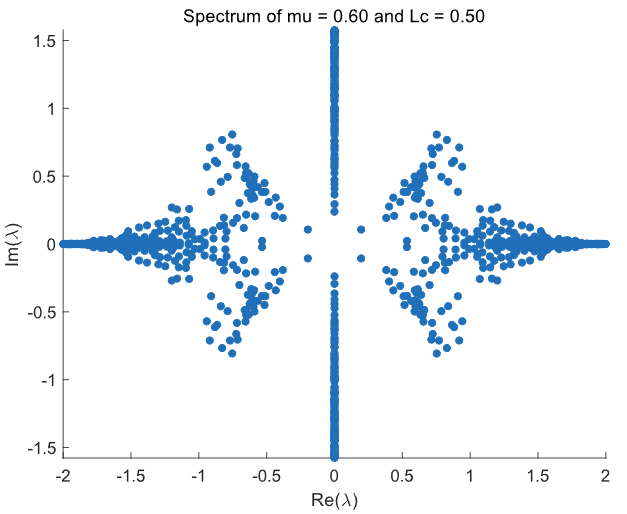}

~~~~~~~~\qquad\quad
$\textbf{(a)}\qquad\qquad\qquad\qquad\qquad\quad\qquad
 \textbf{(b)}\qquad\qquad\qquad\qquad\qquad\quad\qquad\textbf{(c)}$\\
\caption{\small The numerical simulation results of chaotic wave field evolution with the initial condition of a plane wave superimposed by random perturbations, where the correlation length is fixed at \(L_c = 0.5\) and three values \(\mu = 0.1,\,0.3,\,0.6\) are adopted.(a)-(c): Spectra calculated from the initial wave field \(u(x,0)\).}
\label{fig3}
\end{figure}

We have investigated the fundamental behaviors of analytical solutions in the above discussion. Next, we commit to examine the spectral characteristics of the chaotic wave field.
As verified in the second part of this paper, the spectrum is independent of time $t$. Therefore, the initial potential \(u(x,0)\) at \(t=0\) can be adopted to calculate the spectrum. Similar to the spectral computation for analytical solutions, the Fourier collocation method is employed to obtain the spectra under the initial condition Eq. \eqref{19} with different \(\mu\) values, as illustrated in Figs. \ref{fig3}(a)-(c).For \(\mu = 0.1\), it can be seen from Fig. \ref{fig3}(a) that almost all eigenvalues lie on the coordinate axes or on the lines \(\pm\mathrm{Re}(\lambda)=\pm\mathrm{Im}(\lambda)\). As \(\mu\) increases, the eigenvalues gradually deviate and scatter away from the coordinate axes.
Physically, the soliton velocity is governed by the spectral parameter \(\lambda=\alpha+i\beta\). According to Eq. \eqref{8}, the soliton velocity depends on \(\alpha^2-\beta^2\), while the amplitude is mainly determined by the imaginary part of the eigenvalue. Random noise in the chaotic field excites solitons with nonzero velocities, whose corresponding eigenvalues possess both nonvanishing real and imaginary parts and satisfy \(\alpha^2-\beta^2\neq0\). This causes the eigenvalues to disperse off the coordinate axes and generates abundant solitons with nonzero velocity and amplitude in the chaotic wave field, accompanied by an increase in the total number of solitons.

\begin{figure}[H]
\centering
\includegraphics[width=9.0cm,height=7.8cm,angle=0]{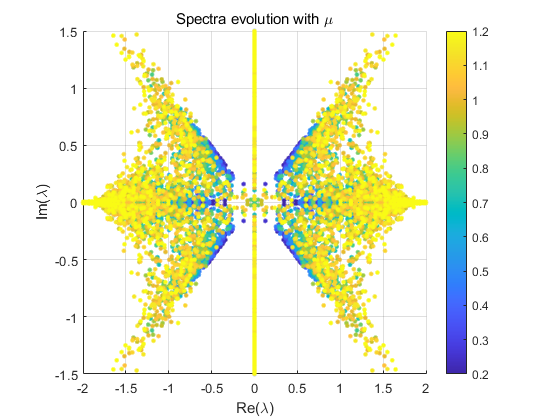}
\caption{\small Evolution of the spectrum with \(\mu\) under fixed \(L_c = 0.5\) and \(\mu\) changing from $0.2$ to $1.2$.}
\label{fig4}
\end{figure}

To intuitively reveal the effect of \(\mu\) on spectral parameters, we plot the spectral variation with \(\mu\) in Fig. \ref{fig4}. It is clearly observed that the spectral parameters become more dispersed as \(\mu\) rises, which is consistent with the theoretical analysis. Since the spectrum (eigenvalues) remains invariant during propagation and is independent of time, the evolution of the chaotic wave field is dominated by the relative proportion of solitons and breathers \cite{15}. An increase in $\mu$ will drive the system to transform from breather turbulence to soliton turbulence. When $\mu$ reaches its maximum value, the breathers completely dissipate, and the chaotic characteristics are entirely dominated by the solitons and their collision effects.

\begin{figure}[H]
 \includegraphics[width=4.8cm,height=4.6cm,angle=0]{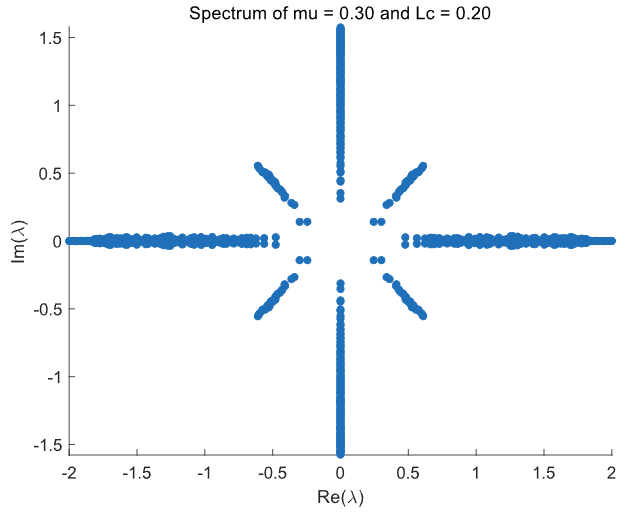}
\includegraphics[width=4.8cm,height=4.6cm,angle=0]{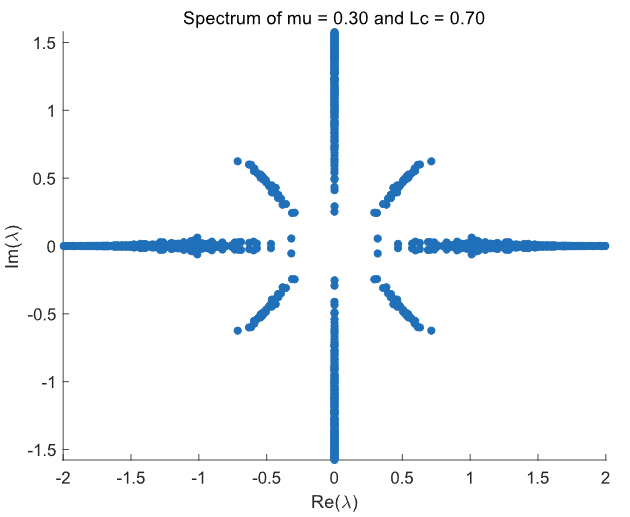}
\includegraphics[width=4.8cm,height=4.6cm,angle=0]{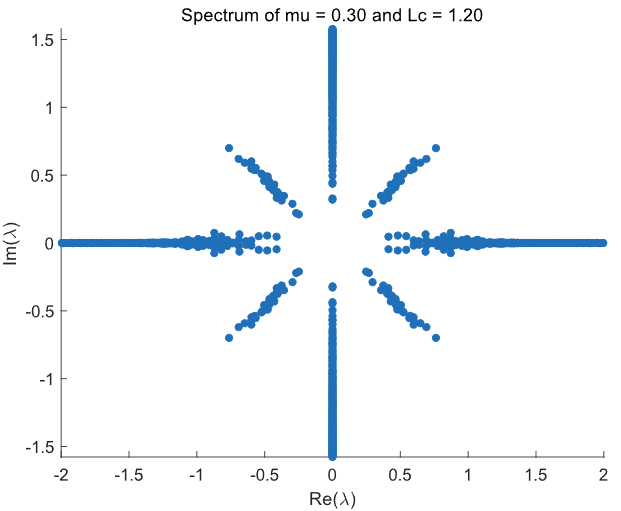}

~~~~~~~~\qquad\quad
$\textbf{(a)}\qquad\qquad\qquad\qquad\qquad\quad\qquad
 \textbf{(b)}\qquad\qquad\qquad\qquad\qquad\quad\qquad\textbf{(c)}$\\
\caption{\small The numerical simulation results of chaotic wave field evolution for a plane wave superimposed with random perturbations as the initial condition, with fixed \(\mu = 0.3\) and three distinct values \(L_c = 0.2,\,0.7,\,1.2\).(a)-(c) Spectra corresponding to the initial wave field \(u(x,0)\).}
\label{fig5}
\end{figure}

\begin{figure}[H]
\centering
\includegraphics[width=9.0cm,height=7.8cm,angle=0]{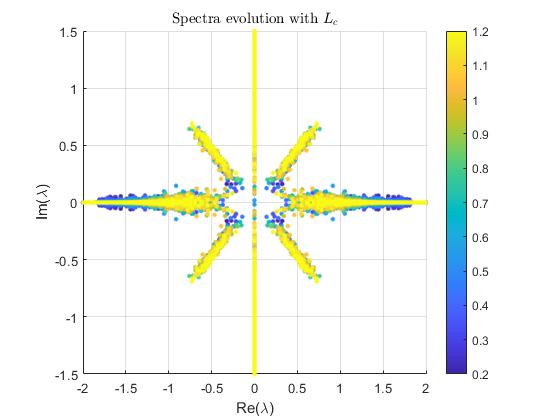}
\caption{\small Evolution of the spectrum with \(L_c\) under fixed \(\mu = 0.3\) and \(L_c\) changing from $0.2$ to $1.2$.}
\label{fig6}
\end{figure}

We further calculate the spectra of the chaotic wave field for different \(L_c\), as displayed in Figs. \ref{fig5}(a)-(c). Unlike the case of \(\mu\), the eigenvalues for all three choices of \(L_c\) are distributed around the coordinate axes. As \(L_c\) increases, the eigenvalues gradually concentrate closer to the coordinate axes and the four diagonal lines, implying a reduction in the number of solitons in the wave field. In other words, \(L_c\) can regulate the quantity of solitons, while its effect on breathers is almost negligible.
Since only three discrete values of \(L_c\) are illustrated in Figs. \ref{fig5}(a)-(c), the underlying trend cannot be clearly identified. To provide a more intuitive presentation, the spectral evolution with continuous variation of \(L_c\) is plotted in Fig. \ref{fig6}. It can be clearly observed that increasing \(L_c\) drives the eigenvalues to approach the real axis, the imaginary axis, and the four lines \(\pm\mathrm{Re}(\lambda)=\pm\mathrm{Im}(\lambda)\), which further demonstrates that a larger \(L_c\) suppresses the number of solitons in the chaotic wave field.

\subsection{Initial field intensity, maximum amplitude and PDF}
\begin{figure}[H]
\includegraphics[width=4.8cm,height=4.6cm,angle=0]{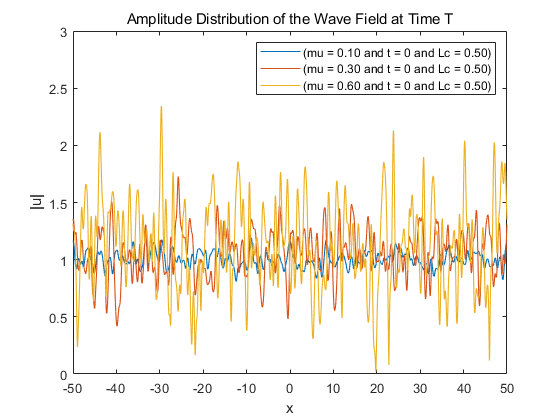}
\includegraphics[width=4.8cm,height=4.6cm,angle=0]{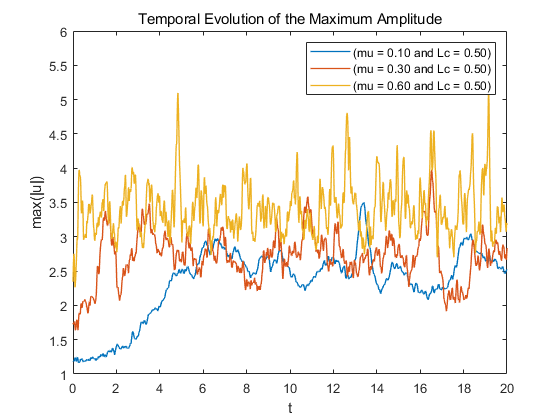}
\includegraphics[width=4.8cm,height=4.6cm,angle=0]{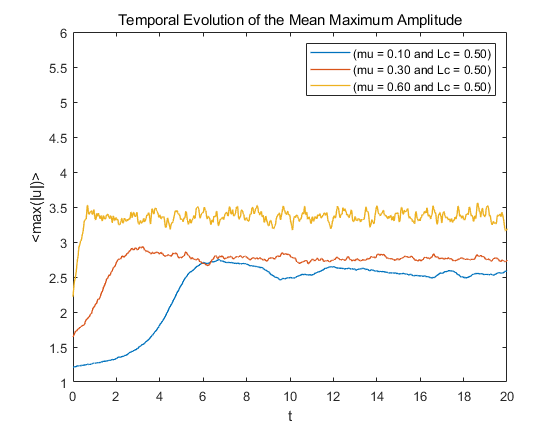}

~~~~~~~~\qquad\quad
$\textbf{(a)}\qquad\qquad\qquad\qquad\qquad\quad\qquad
 \textbf{(b)}\qquad\qquad\qquad\qquad\qquad\quad\qquad\textbf{(c)}$\\
\caption{\small The numerical simulation of chaotic wave field evolution under the initial condition with random perturbations, where \(L_c = 0.5\) and \(\mu = 0.1\) (blue line), \(\mu = 0.3\) (red line), and \(\mu = 0.6\) (yellow line).(a) Intensity distribution of the initial wave field \(u(x,0)\);(b) Temporal evolution of the maximum wave-field amplitude;(c) Evolution of the mean highest amplitude.}
\label{fig7}
\end{figure}

In this part, we adopt the SSF method \cite{jia2} to investigate the numerical evolution of Eq. \ref{3} with a plane wave initial condition containing random noise, including intensity distribution of the initial wave field $u(x,0)$, temporal evolution of the maximum wave-field amplitude and the probability density function (PDF) of field strength. Fig. \ref{fig7}(a) illustrates the influence of varying values of \(\mu\) on the intensity distribution of \(u(x, 0)\) under a fixed correlation length \(L_c\). It can be clearly observed that when \(\mu = 0.1\), the amplitude fluctuates randomly around 1 with a relatively small random deviation. In contrast, for \(\mu = 0.3\) and \(\mu = 0.6\), the chaotic deviations around the continuous wave increase significantly. This indicates that when the correlation length \(L_c\) is fixed, the chaotic deviation of the initial condition \(u(x, 0)\) is positively correlated with \(\mu\). As we all know, initial random conditions can lead to modulation instability, which in turn evolves into a chaotic wave field \cite{15,21}. We calculated the evolution of the maximum amplitude of the simulated chaotic wave field over time under the above initial conditions, and the results are shown in Fig. \ref{fig7}(b). According to the modulation instability theory, for \(\mu = 0.1\), the maximum amplitude increases exponentially in the early stage of evolution. After evolving for a period of time along the time axis, this value converges to a chaotic state, i.e., the maximum amplitude oscillates back and forth within a certain interval with rapid changes.
The larger the value of \(\mu\), the faster it converges to the chaotic state, and the larger the maximum amplitude of the wave field after reaching the chaotic state. The evolution processes of \(\mu = 0.3\) and \(\mu = 0.6\) are similar, with the difference that the maximum amplitude of the wave field corresponding to \(\mu = 0.6\) is significantly larger than that of \(\mu = 0.3\) after reaching the chaotic state. This indicates that the greater the intensity of the initial condition (which can also be understood as higher energy), the shorter the time required for the chaotic wave field to converge to the chaotic state, and the higher the peak amplitude of the chaotic state. The convergence process of the system towards the turbulent limit state can be further illustrated by Fig. \ref{fig7}(c), which shows the evolution law of the average peak amplitude. Each curve is the average of the peak amplitudes obtained from one hundred random experiments. It is clearly visible that the steady-state value of the average peak amplitude is all higher than 2.5. From this, we can also see that as $\mu$ increases, the maximum amplitude increases. Moreover, it is not difficult to see from the figure that after the system reaches the chaotic limit state at a certain position, it will basically remain unchanged and continuously maintain this state along the $t$-axis, and will not return to the initial state at any evolving position.

\begin{figure}[H]
\includegraphics[width=4.8cm,height=4.6cm,angle=0]{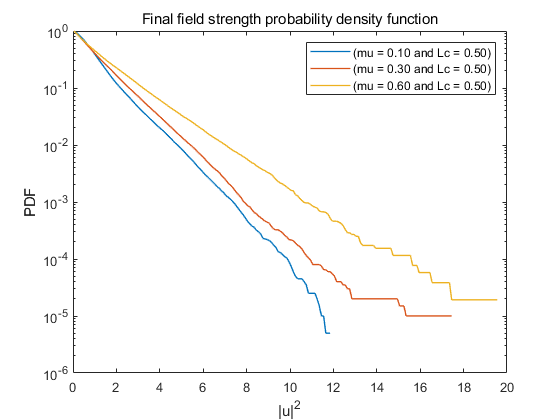}
\includegraphics[width=4.8cm,height=4.6cm,angle=0]{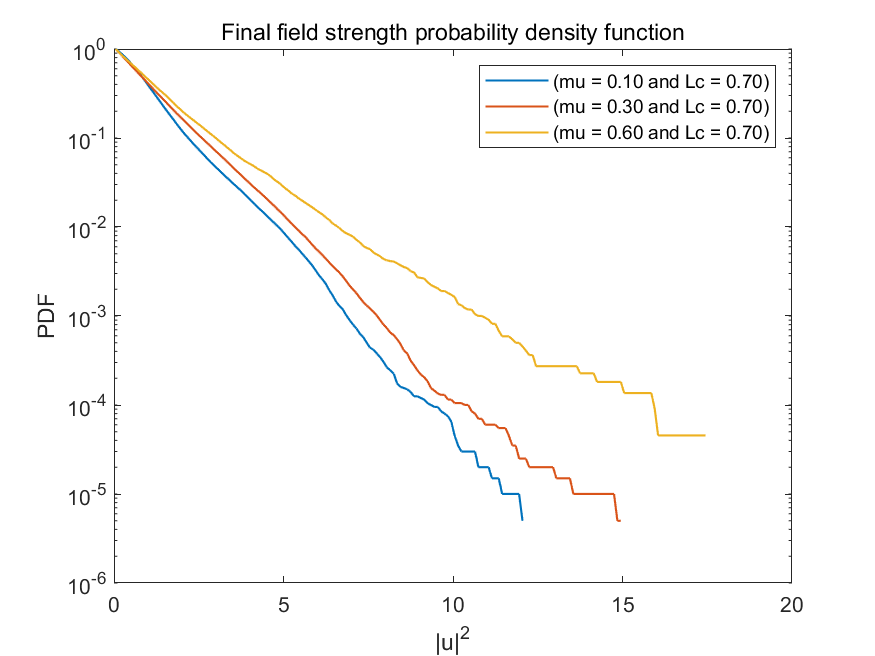}
\includegraphics[width=4.8cm,height=4.6cm,angle=0]{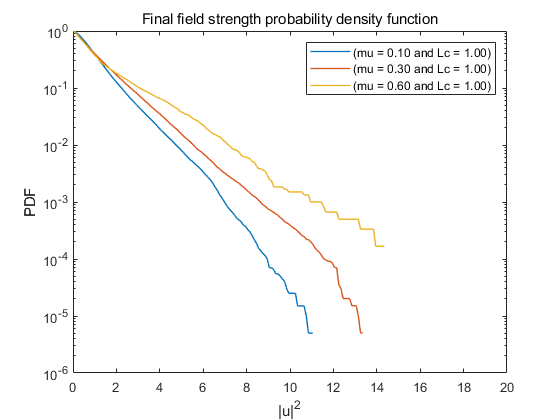}

~~~~~~~~\qquad\quad
$\textbf{(a)}\qquad\qquad\qquad\qquad\qquad\quad\qquad
 \textbf{(b)}\qquad\qquad\qquad\qquad\qquad\quad\qquad\textbf{(c)}$\\
\caption{\small Probability density functions calculated
for three different values of \(\mu\) and for (a) \(L_c\)=0.5, (b) \(L_c\)=0.7,
and (c) \(L_c\)=1.0.}
\label{fig11}
\end{figure}

In addition, we also calculated the PDF of field strength, as it is closely related to the probability of rogue wave occurrence in the chaotic wave field \cite{6,42}. Fig. \ref{fig11} shows the field intensity PDF obtained after evolving 100 units along the $t$-axis. Obviously, the larger the initial \(\mu\) value, the higher the tail of the corresponding PDF. Therefore, it can be predicted that when the wave field converges to the chaotic state, the probability of rogue wave generation will increase significantly with the increase of \(\mu\). An increase in \(\mu\) will generate solitary waves with higher amplitudes and velocities, thereby increasing the field strength at the collision positions of the solitons and also increasing the frequency of collisions.

\begin{figure}[H]
\includegraphics[width=4.8cm,height=4.6cm,angle=0]{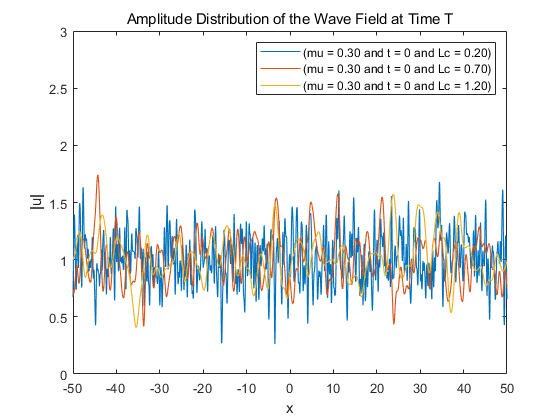}
\includegraphics[width=4.8cm,height=4.6cm,angle=0]{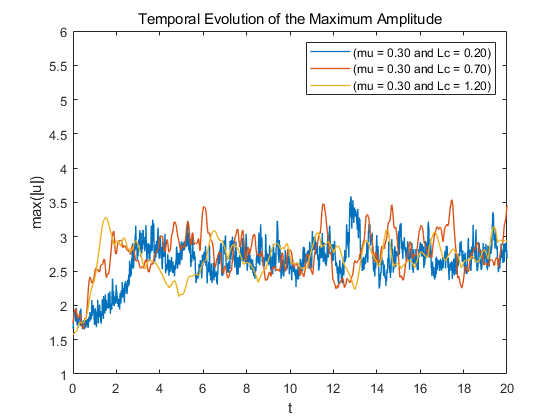}
\includegraphics[width=4.8cm,height=4.6cm,angle=0]{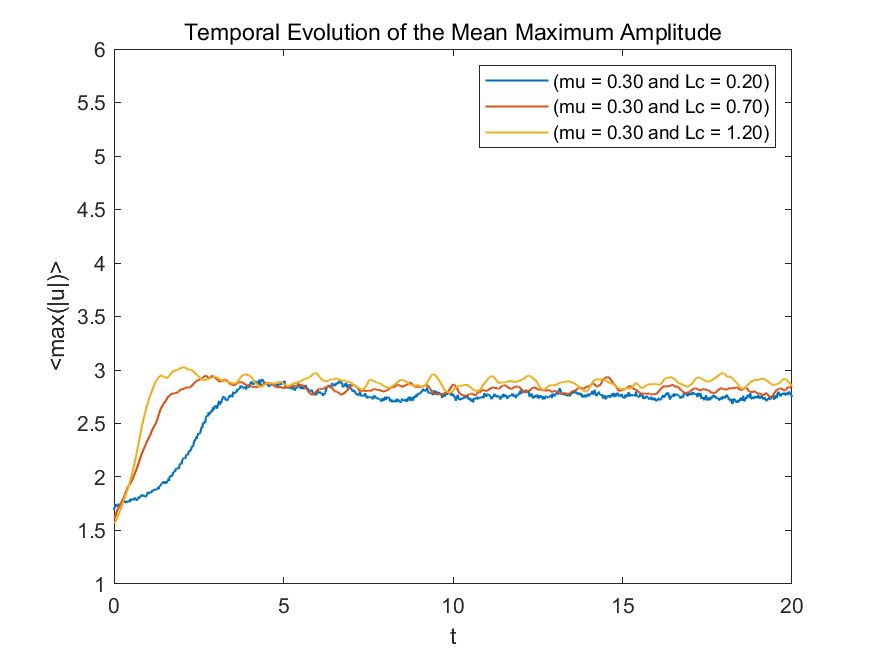}

~~~~~~~~\qquad\quad
$\textbf{(a)}\qquad\qquad\qquad\qquad\qquad\quad\qquad
 \textbf{(b)}\qquad\qquad\qquad\qquad\qquad\quad\qquad\textbf{(c)}$\\
\caption{\small The numerical simulation of chaotic wave field evolution under the initial condition with random perturbations, where \(\mu=0.3\), \(L_c=0.2\) (blue line), \(L_c=0.7\) (red line), and \(L_c=1.2\) (yellow line).(a) Intensity distribution of the initial wave field \(u(x,0)\);(b) Temporal evolution of the maximum amplitude of the wave field;(c) Evolution of the mean highest amplitude.}
\label{fig8}
\end{figure}

\begin{figure}[H]
\includegraphics[width=4.8cm,height=4.6cm,angle=0]{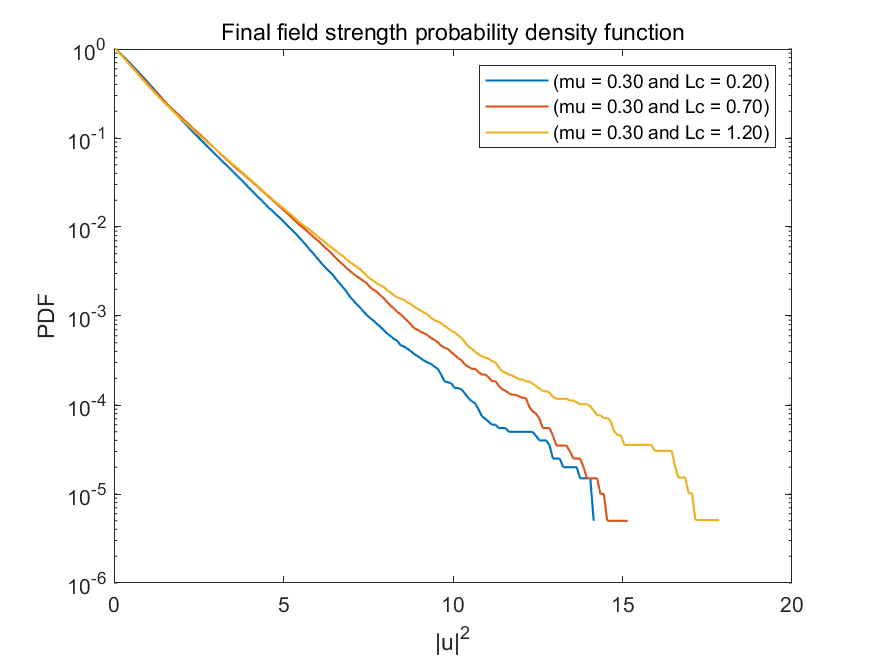}
\includegraphics[width=4.8cm,height=4.6cm,angle=0]{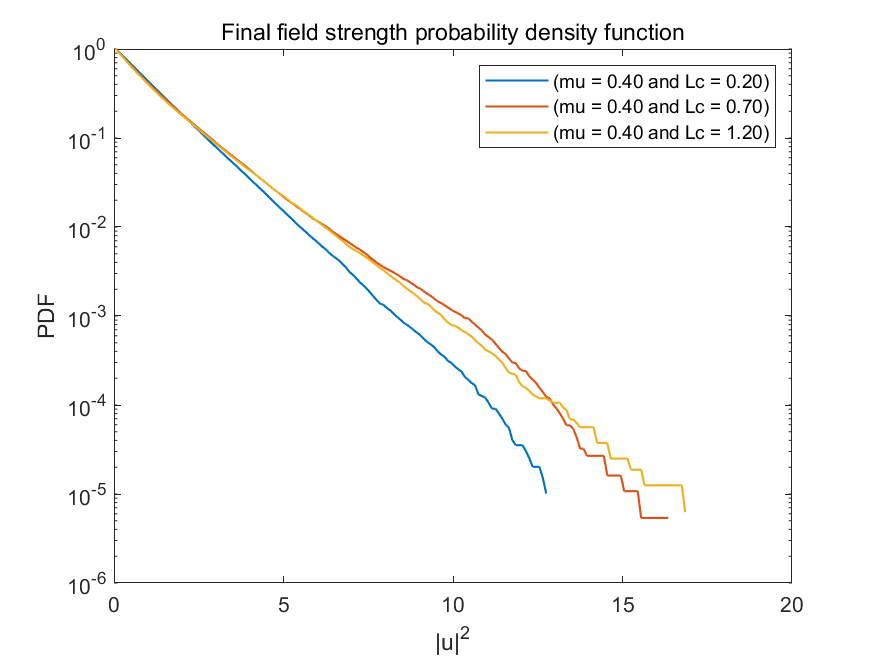}
\includegraphics[width=4.8cm,height=4.6cm,angle=0]{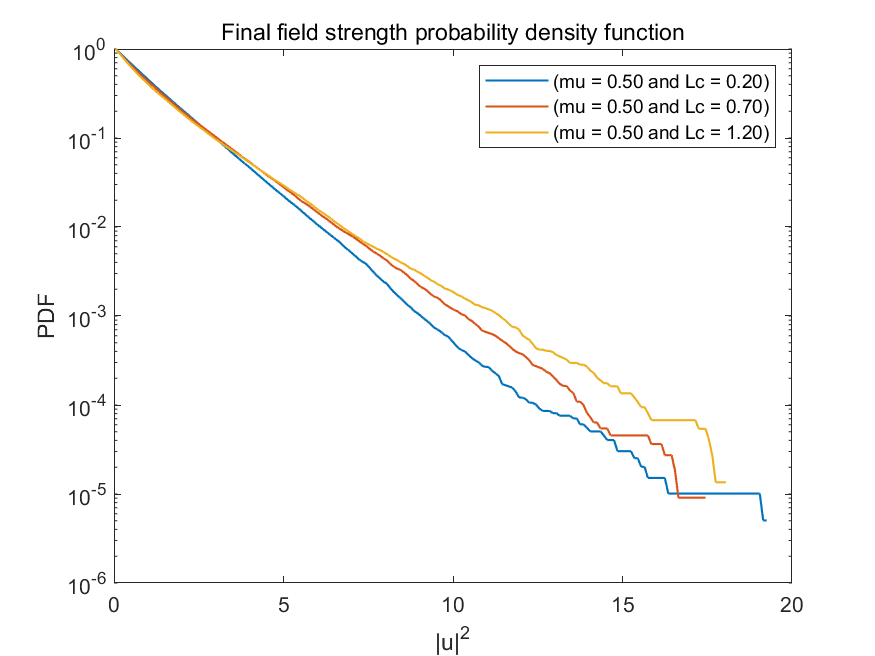}

~~~~~~~~\qquad\quad
$\textbf{(a)}\qquad\qquad\qquad\qquad\qquad\quad\qquad
 \textbf{(b)}\qquad\qquad\qquad\qquad\qquad\quad\qquad\textbf{(c)}$\\

\caption{\small Probability density functions calculated
for three different values of \(L_c\) and for (a) \(\mu\)=0.3, (b) \(\mu\)=0.4,
and (c) \(\mu\)=0.5.}
\label{fig12}
\end{figure}

Next, we adopt the same numerical method to analyze the influence of the correlation length \(L_c\) on the evolution of the chaotic wave field, where \(L_c\) is closely related to the oscillation frequency of the initial condition. As shown in Fig. \ref{fig8}(a), the oscillation frequency decreases as \(L_c\) increases, while the amplitude remains almost unchanged.
In Fig. \ref{fig8}(b), a larger \(L_c\) enables the wave field to converge to the chaotic state more rapidly. After reaching chaos, the fluctuation range of the maximum amplitude is roughly identical for different \(L_c\), indicating that \(L_c\) hardly affects the peak amplitude in the chaotic regime. Nevertheless, it is obvious that a smaller \(L_c\) leads to a high oscillation frequency of the maximum amplitude even in the chaotic state, whereas the oscillation frequency declines remarkably as \(L_c\) increases. This reveals that \(L_c\) also modulates the overall oscillation frequency of the wave field after entering the chaotic state. Fig.\ref{fig8}(c) illustrates the evolution of the average peak amplitude under different correlation lengths \(L_c\). Each curve is plotted by averaging the peak amplitudes obtained from one hundred groups of random numerical experiments. It can be observed that the maximum field amplitude remains nearly unchanged as the parameter \(L_c\) increases.
Fig. \ref{fig8}(c) also indicates that once the system reaches a chaotic limiting state at a certain evolutionary stage, its amplitude will stay almost invariant along the time axis $t$. Fig. \ref{fig12} presents the PDF of the field intensity after propagating 100 units along the $t$-axis. Different from the behavior of \(\mu\) in Fig. \ref{fig11}, the variation of \(L_c\) has a weak influence on the PDF, and the increase in \(L_c\) has no significant effect on the tail lift. Accordingly, it can be inferred that \(L_c\) does not significantly affect the generation probability of rogue waves once the wave field evolves into chaos.

\subsection{Evolution of chaotic patterns}
\begin{figure}[H]
\includegraphics[width=4.8cm,height=4.6cm,angle=0]{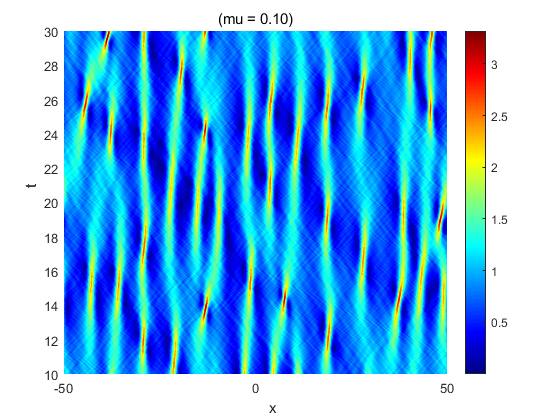}
\includegraphics[width=4.8cm,height=4.6cm,angle=0]{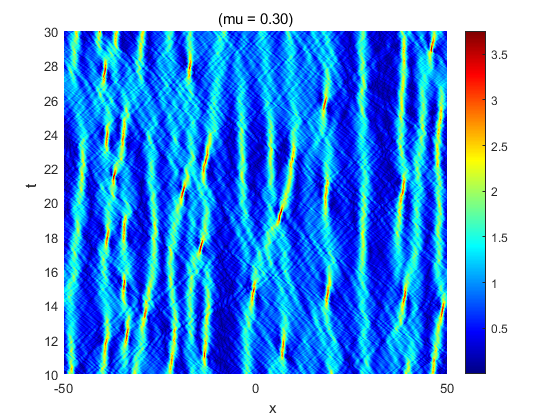}
\includegraphics[width=4.8cm,height=4.6cm,angle=0]{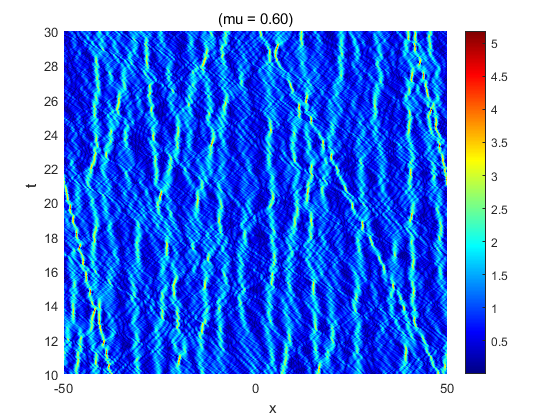}

~~~~~~~~\qquad\quad
$\textbf{(a1)}\qquad\qquad\qquad\qquad\qquad\qquad
 \textbf{(a2)}\qquad\qquad\qquad\qquad\qquad\qquad\textbf{(a3)}$\\

 \includegraphics[width=4.8cm,height=4.6cm,angle=0]{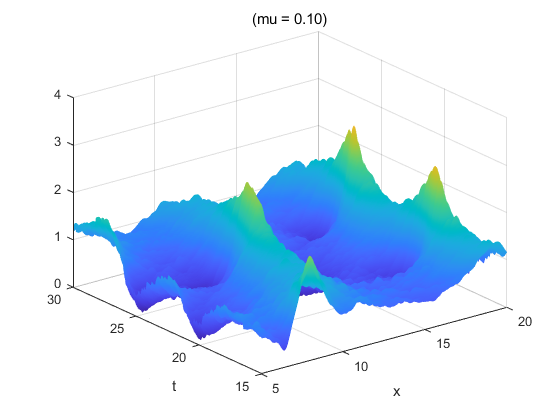}
\includegraphics[width=4.8cm,height=4.6cm,angle=0]{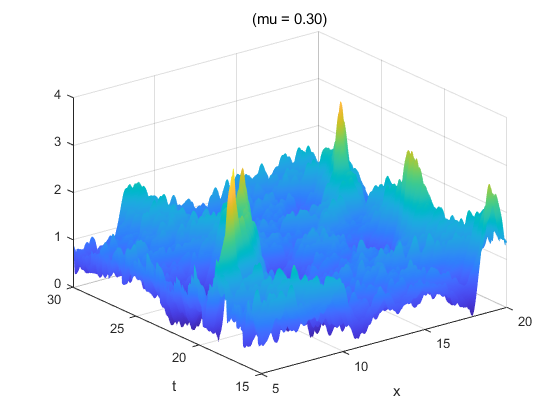}
\includegraphics[width=4.8cm,height=4.6cm,angle=0]{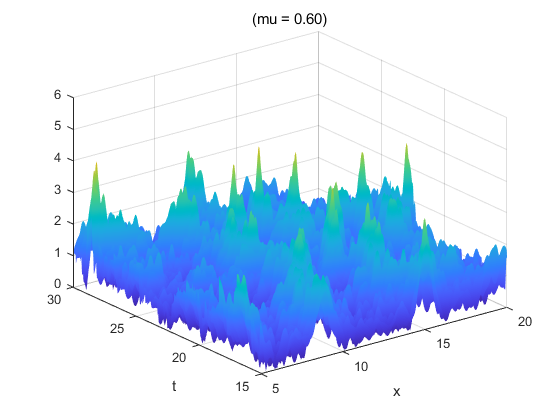}

~~~~~~~~\qquad\quad
$\textbf{(b1)}\qquad\qquad\qquad\qquad\qquad\qquad
 \textbf{(b2)}\qquad\qquad\qquad\qquad\qquad\qquad\textbf{(b3)}$\\
\caption{\small The numerical simulation results of chaotic wave field evolution with the initial condition of a plane wave superimposed by random perturbations, where the correlation length is fixed at \(L_c = 0.5\) and three values \(\mu = 0.1,\,0.3,\,0.6\) are adopted.[(a1)-(a3)] Two-dimensional spatiotemporal projections of the field intensity \(|u|\) at point \((x,t)\);[(b1)-(b3)] Three-dimensional views of the field intensity \(|u|\).}
\label{fig9}
\end{figure}

To better understand the transformation process of the wave field components from the breather turbulence to the soliton turbulence, we conducted numerical simulations of the evolution of the chaotic pattern. Figs. \ref{fig9}(a1)-(a3) respectively show the field amplitudes $\vert u\vert$ when $\mu=0.1, 0.3$, and $0.6$. Fig. \ref{fig9}(a1) show that breather waves are excited and interact with each other within the wave field. The case \(\mu=0.3\) corresponds to an intermediate state, as shown in Fig. \ref{fig9}(a2), where mutual interactions occur among breathers, solitons, and background radiation. For Fig. \ref{fig9}(a3), the chaotic wave field is dominated by solitons.
This indicates that increasing the random parameter \(\mu\) drives the transition of the chaotic field from breather turbulence to soliton turbulence. Consistent with the spectral results in Figs. \ref{fig3}(a)-(c), the number of solitons grows evidently with increasing \(\mu\).
In addition, the amplitude scales of the wave fields are displayed on the right side of Figs. \ref{fig9}(a1)-(a3). The peak amplitudes are 3, 3.5, and 5 for \(\mu=0.1\), \(\mu=0.3\), and \(\mu=0.6\), respectively. This demonstrates that the peak amplitude increases with \(\mu\), which correspondingly raises the probability of rogue wave occurrence. Figs. \ref{fig9}(b1)-(b3) present the extracted three-dimensional snapshots selected from the spatiotemporal evolution diagrams in Figs. \ref{fig9}(a1)-(a3).

\begin{figure}[H]
\includegraphics[width=4.8cm,height=4.6cm,angle=0]{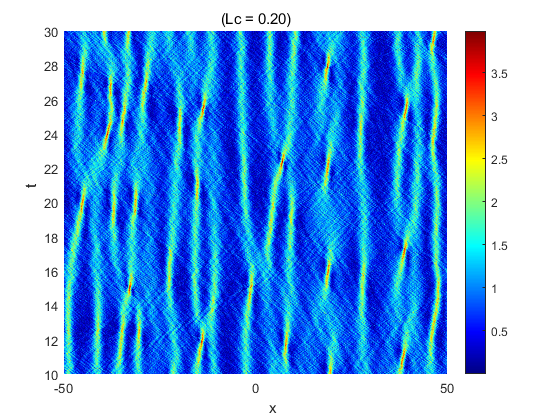}
\includegraphics[width=4.8cm,height=4.6cm,angle=0]{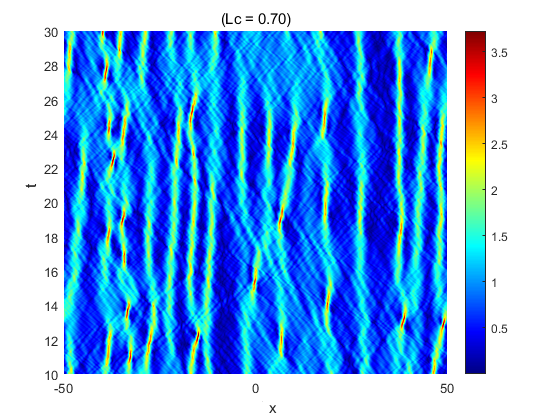}
\includegraphics[width=4.8cm,height=4.6cm,angle=0]{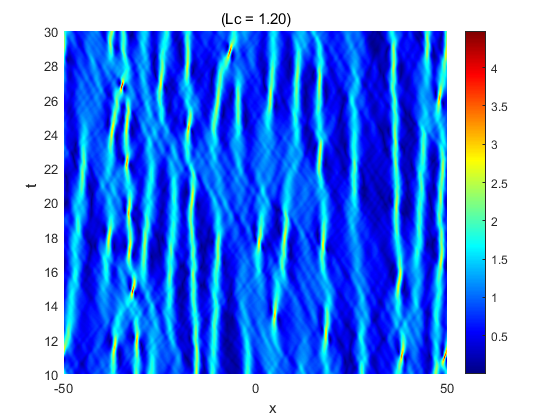}

~~~~~~~~\qquad\quad
$\textbf{(a1)}\qquad\qquad\qquad\qquad\qquad\qquad
 \textbf{(a2)}\qquad\qquad\qquad\qquad\qquad\qquad\textbf{(a3)}$\\

 \includegraphics[width=4.8cm,height=4.6cm,angle=0]{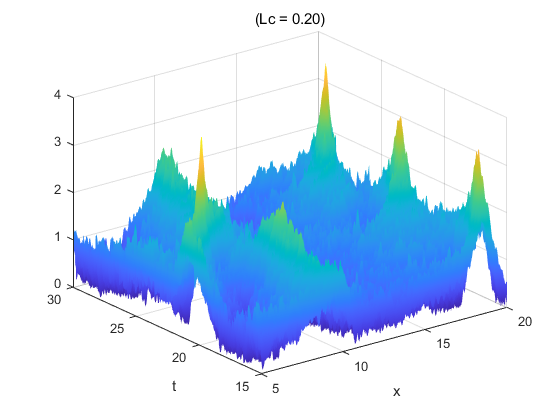}
\includegraphics[width=4.8cm,height=4.6cm,angle=0]{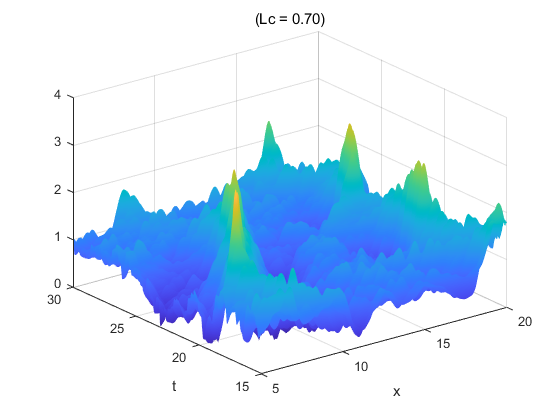}
\includegraphics[width=4.8cm,height=4.6cm,angle=0]{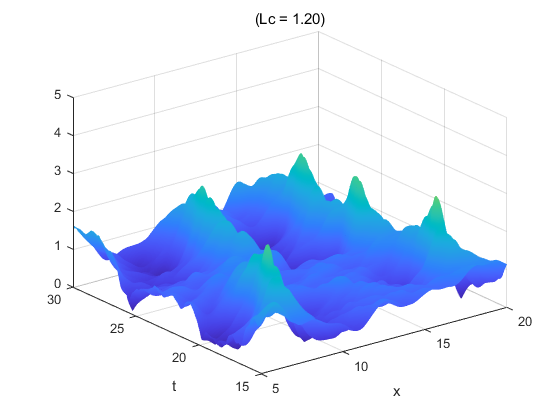}

~~~~~~~~\qquad\quad
$\textbf{(b1)}\qquad\qquad\qquad\qquad\qquad\qquad
 \textbf{(b2)}\qquad\qquad\qquad\qquad\qquad\qquad\textbf{(b3)}$\\
\caption{\small The numerical simulation results of chaotic wave field evolution for a plane wave superimposed with random perturbations as the initial condition, with fixed \(\mu = 0.3\) and three distinct values \(L_c = 0.2,\,0.7,\,1.2\).[(a1)-(a3)] Two-dimensional spatiotemporal projections representing the field intensity \(|u|\) at the point \((x,t)\);[(b1)-(b3)] Three-dimensional views of the field intensity \(|u|\).}
\label{fig10}
\end{figure}

In order to further analyze the effect of parameter \(L_c\) on the evolution of the chaotic pattern, we conducted numerical simulations of the chaotic patterns under different \(L_c\) values. Figs. \ref{fig10}(a1)-(a3) illustrate the breathers, solitons, large-amplitude waves, background radiation and interact with each other in the wave field. Unlike the regulatory effect of \(\mu\), no transition from breather turbulence to soliton turbulence is observed with the increase of \(L_c\), which is consistent with the spectral results. Meanwhile, the peak amplitude of the wave field does not grow with increasing \(L_c\), and verifies that \(L_c\) has a minor impact on the probability of rogue wave occurrence. Figs. \ref{fig10}(b1)-(b3) show the extracted three-dimensional snapshots of the wave field corresponding to Figs. \ref{fig10}(a1)-(a3).

\subsection{Wave action spectrum for chaotic wave fields}
\begin{figure}[H]
\includegraphics[width=4.8cm,height=4.6cm,angle=0]{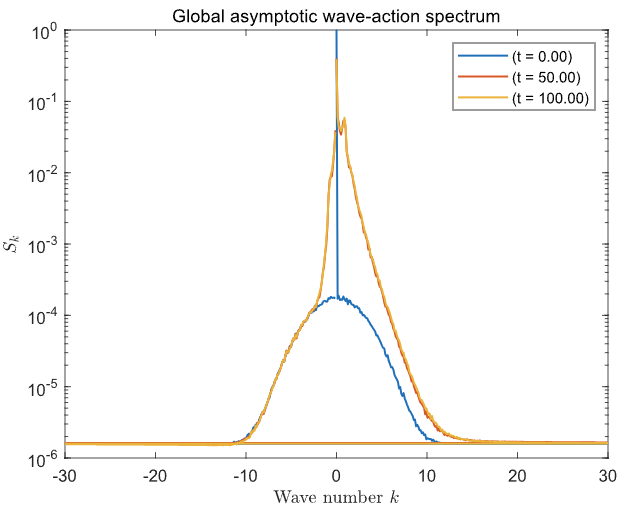}
\includegraphics[width=4.8cm,height=4.6cm,angle=0]{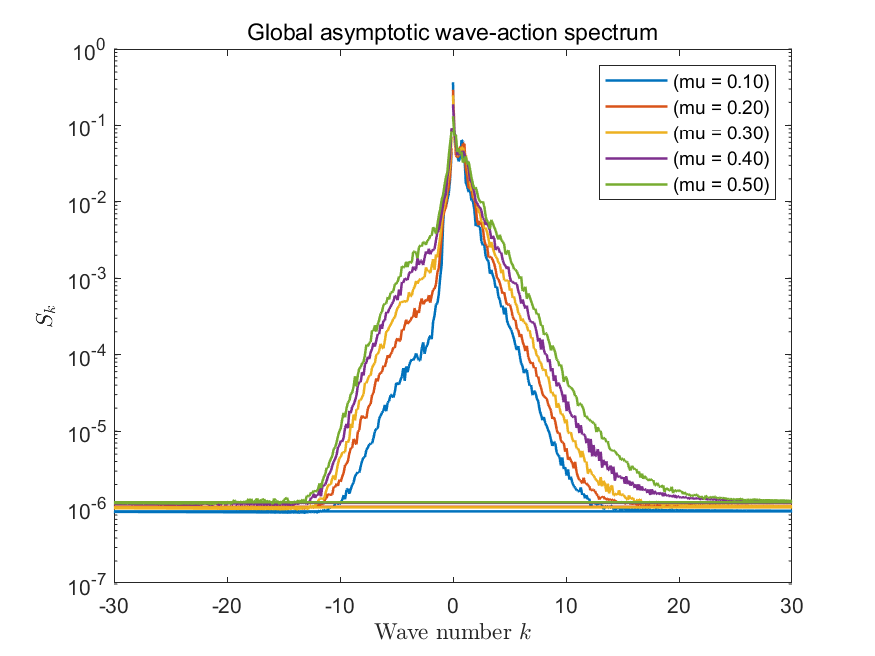}
\includegraphics[width=4.8cm,height=4.6cm,angle=0]{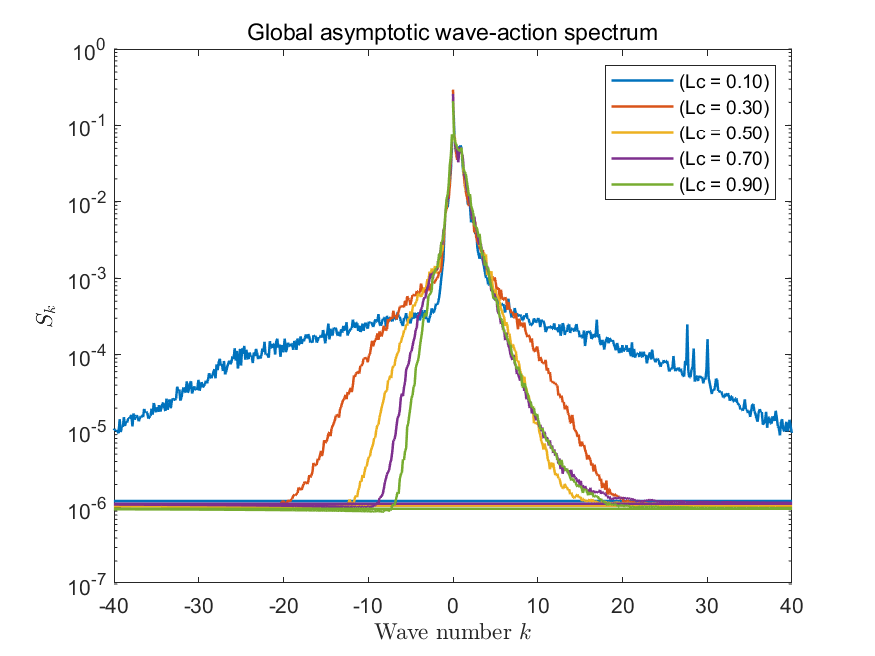}

~~~~~~~~\qquad\quad
$\textbf{(a)}\qquad\qquad\qquad\qquad\qquad\quad\qquad
 \textbf{(b)}\qquad\qquad\qquad\qquad\qquad\quad\qquad\textbf{(c)}$\\
\caption{\small Wave action spectrum.(a) Ensemble-averaged wave action spectra over 500 random trials at different moments (\(t=0\) (blue line), \(t=50\) (red line), \(t=100\) (yellow line)) under fixed parameters \(\mu=0.1\) and \(L_c=0.5\).(b) Wave action spectra for different \(\mu\) values at \(t=50\) during chaotic wave field evolution (ensemble-averaged over 100 random trials).(c) Wave action spectra for different \(L_c\) values at \(t=50\) during chaotic wave field evolution (ensemble-averaged over 100 random trials).}
\label{fig13}
\end{figure}

In this part, we further examine the wave action spectrum of the GI equation. The wave-action spectrum, generally denoted as \(S_k(t)\), is one of the essential characteristic quantities of chaotic wave fields. In general, when the characteristic parameters of the chaotic wave field remain steady, the wave-action spectrum presents a relatively stable profile.
The wave-action spectrum is conventionally defined as \cite{42,43}
\begin{align}\label{21}
S_k(t) = \big\langle\,|\psi_k(t)|^2\,\big\rangle,
\end{align}
where \(\langle\,\cdot\,\rangle\) represents the ensemble average over multiple random realizations, and \(\psi_k(t)\) denotes the semi-discrete Fourier transform
\begin{align}\label{22}
&\psi_k(t) = \mathcal{F}\big[\psi(x,t)\big]
= \frac{1}{L}\int_{-L/2}^{L/2} \psi(x,t)e^{-ikx}\mathrm{d}x,\notag\\  &\psi(x,t) = \mathcal{F}^{-1}\big[\psi_k(t)\big]
= \sum_{k}\psi_k(t)e^{ikx}.
\end{align}
To numerically compute the wave-action spectrum in MATLAB, Eq. \eqref{22} is discretized into the following form
\begin{align}\label{23}
\psi_k(t) = \frac{1}{L}\sum_{i}\psi(x_i,t)\mathrm{d}x.
\end{align}
Here, \(\psi(x_i,t)\) represents the wave potential at discrete spatial point \(x_i\) and time $t$, which corresponds to \(u(x_i,t)\) for the GI equation.
Substituting Eq. \eqref{23} into Eq. \eqref{21}, we can evaluate \(S_k(t)\). A series of random numerical experiments are carried out, and the averaged wave-action spectrum over independent random realizations is presented in Fig. \ref{fig13}.
Fig. \ref{fig13}(a) shows the ensemble-averaged results obtained from 500 random trials at three typical evolutionary moments \(t=0,\,50,\,100\). It can be clearly observed that the wave-action spectra at \(t=50\) and \(t=100\) almost completely overlap, indicating that the chaotic wave field rapidly relaxes to a statistically stationary spectral state. Moreover, an interesting feature is found that the wave-action spectrum of the GI equation exhibits an asymmetric distribution during temporal evolution.
To further explore the influence of random parameters on the wave-action spectrum, we plot the spectral profiles at \(t=50\)(a time instant at which the spectrum has reached a steady state according to Fig. \ref{fig13}(a)) under different parameter conditions, as shown in Figs. \ref{fig13}(b) and (c).
As illustrated in Fig. \ref{fig13}(b), the overall magnitude of the wave-action spectrum increases with \(\mu\), while an opposite linear trend appears at \(k=0\), where the spectral value decreases as \(\mu\) grows.
From Fig. \ref{fig13}(c), a negative correlation is observed between the wave-action spectrum and the correlation length \(L_c\): a larger \(L_c\) corresponds to a weaker wave-action spectrum. Specifically, the spectral distribution covers a broad range for \(L_c=0.1\), whereas it becomes much narrower for \(L_c=0.9\).

\section{Conclusion}
This paper investigates the characteristics of integrable turbulence and chaotic wave fields governed by the GI equation. First, to establish the correspondence between spectra and different analytical solutions, we numerically calculate the spectral profiles of soliton, breather, and rogue wave solutions via the Fourier collocation method. It is found that the spectral profiles of the GI equation exhibit a special double symmetry with respect to both the real and imaginary axes.
Subsequently, we simulate the chaotic wave field under the initial condition of a plane wave with random noise. The variations of the initial field intensity and maximum field amplitude with the standard deviation \(\mu\) reveal that a larger standard deviation leads to more pronounced chaotic deviation around the continuous wave, a faster convergence to the chaotic state, and a higher peak amplitude of the wave field. Furthermore, the height of the PDF tail rises significantly as \(\mu\) increases, indicating a substantial growth in the generation probability of rogue waves. Numerical results also demonstrate that increasing \(\mu\) triggers a transition of turbulence type from breather-dominated turbulence to soliton-dominated turbulence.
In addition, we explore the influence of the correlation length \(L_c\) on the evolution of the plane wave with random initial noise. The results show that \(L_c\) mainly affects the convergence rate of random initial fields toward the chaotic state, while exerting negligible impacts on the maximum amplitude and the generation probability of rogue waves. Notably, an increase in \(L_c\) correspondingly reduces the number of solitons in the chaotic wave field.
Finally, we investigate the properties and phenomena associated with the wave-action spectrum. In chaotic wave fields, the wave-action spectrum generally reaches a steady state rapidly and remains nearly time-independent thereafter. We further examine the evolution of the wave-action spectrum under different random parameters \(\mu\) and \(L_c\). The wave-action spectrum at a fixed wave-number $k$ increases with rising \(\mu\), showing a positive correlation. In contrast, \(L_c\) presents an opposite trend: the wave-action spectrum at the same wave-number $k$ decreases as \(L_c\) increases, exhibiting a negative correlation. This study provides a theoretical reference for understanding the generation mechanisms of integrable turbulence and rogue waves in the GI equation from numerical and statistical perspectives.

In the field of integrable turbulence, there are still many research directions that are worthy of further exploration and in-depth study:
First, the SSF method employed in this study performed exceptionally well under moderate nonlinear intensity conditions. However, when the nonlinear term dominates (e.g., large-amplitude initial perturbations), the selection of time step and spatial discretization accuracy will affect the statistical results of long-term evolution. Future research can introduce adaptive step-size control or high-order splitting schemes (such as the fourth-order Runge-Kutta SSF scheme) to improve the structure-preserving performance of long-time numerical simulations.
Second, the current model is restricted to the one-dimensional framework. For practical physical problems, turbulent behaviors in two-dimensional integrable systems are expected to be more abundant and complex. Future research could focus on exploring two-dimensional or coupled integrable systems, including turbulence caused by vortex-soliton interactions, as well as the distribution and evolution laws of the two-dimensional wave action spectrum. This will entail higher computational costs, but it is expected to uncover entirely new statistical patterns.

\section*{Acknowledgements}

W. Q. Peng is supported by the Fundamental Research Funds for the Central Universities
No. 202513017, National Natural Science Foundation of China No. 12501335, Natural Science
Foundation of Shandong Province No. ZR2025QC1464 and Qingdao Postdoctoral Science Foundation No.
QDBSH20250102147. S. F. Tian is supported by the National Natural Science Foundation of China under Grant No. 12371255, the
Fundamental Research Funds for the Central Universities of CUMT under Grant No. 2024ZDPYJQ1003.

%

\setcounter{equation}{0}
\renewcommand\theequation{A.\arabic{equation}}

\appendix
\section{Appendix: Numerical Method for the GI Eq.\eqref{3}}
\renewcommand{\theequation}{\Alph{section}.\arabic{equation}}
In this work, the SSF method is mainly adopted for numerical simulations. The fundamental idea of the SSF method is to decompose the original governing equation into two subproblems corresponding to the linear and nonlinear terms . For Eq. \eqref{3}, its linear part reads
\begin{align}\label{24}
u_t = i u_{xx}
\end{align}
and the nonlinear part is given by
\begin{align}\label{25}
u_t = u^2 u_x^* + \tfrac12 i u^3 u^{*2}.
\end{align}
We assume that the grid points in Fourier space are defined as \(X_j = \dfrac{2\pi j}{N},\quad j=0,1,2,\dots,N.\)
The numerical approximation of \(u(X_j,t)\) is denoted as \(Q_j(t)\). First, the discrete Fourier transform is employed to solve the linear subproblem.
Let the Fourier transform of $u$ with respect to $x$ be
\begin{align}\label{26}
\mathcal{F}(u) = U,
\end{align}
then the following relations hold:
\begin{align}\label{27}
\mathcal{F}(u_t) = U_t,\quad \mathcal{F}(u_x) = ikU,\quad \mathcal{F}(u_{xx}) = -k^2 U.
\end{align}
Substituting Eq. \eqref{27} into Eq. \eqref{24} yields
\begin{align}\label{28}
U_t = -ik^2 U
\end{align}
whose analytical solution is
\begin{align}\label{29}
U = U_0 \exp\left(-ik^2 t\right).
\end{align}
Discretizing in time (where \(\Delta t\) denotes the time step), we obtain
\begin{align}\label{30}
U^{m+1} = U^m \exp\left(-ik^2 \Delta t\right).
\end{align}
Since
\begin{align}\label{31}
u = \mathcal{F}^{-1}(U),
\end{align}
the time iteration formula can be derived as
\begin{align}\label{32}
Q_j^{m+1} = \mathcal{F}_j^{-1}\Big[\mathcal{F}\big(Q_j^m\big) \exp\big(-ik^2\Delta t\big)\Big],
\end{align}
where \(\mathcal{F}\) and \(\mathcal{F}^{-1}\) represent the Fourier transform and inverse Fourier transform, respectively. \(Q_j^m\) stands for the numerical approximation of \(u(X_j, m\Delta t)\).

From Eq. \eqref{27} , we further derive
\begin{align}\label{33}
u_x = \mathcal{F}^{-1}(ikU) = \mathcal{F}^{-1}\big[ik\mathcal{F}(u)\big].
\end{align}
By discretizing Eq. \eqref{33} and substituting it into Eq. \eqref{25}, the spatially discrete form of the nonlinear part is obtained as
\begin{align}\label{34}
\dot{Q}_j = Q_j^2 \mathcal{F}^{-1}\big[ik\mathcal{F}(Q_j^*)\big] + \frac12 i Q_j^3 {Q_j^*}^2,
\end{align}
where the dot over $Q$ denotes the time derivative. For the time integration of Eq. \eqref{32}, the fourth-order Runge-Kutta method is adopted.

\end{document}